\documentclass{article}
\usepackage{epsfig}
\usepackage{amssymb}

\newcommand{\be}{\begin{equation}}
\newcommand{\ee}{\end{equation}}
\newcommand{\ba}{\begin{eqnarray}}
\newcommand{\ea}{\end{eqnarray}}
\input{tcilatex}

\begin{document}

\title{{\bf Two-dimensional delta potential wells and condensed-matter
physics}}
\author{M. de Llano$^{1}$, A. Salazar$^{2}$ and M.A. Sol\'{\i}s$^{2,3}$ \\
%EndAName
$^{1}$Instituto de Investigaciones en Materiales, UNAM, \\
Apdo. Postal 70-360, 04510 M\'{e}xico, DF, Mexico\\
$^{2}$Instituto de F\'{\i}sica, UNAM, Apdo. Postal 20-364, \\
01000 M\'{e}xico, DF, Mexico \\
$^{3}$Department of Physics, Washington University, \\
St. Louis, Missouri 63130, USA}
\maketitle

\begin{abstract}
It is well-known that a delta potential well in 1D has only one bound state
but that in 3D it supports an {\it infinite} number of bound states with 
{\it infinite} binding energy for the lowest level. We show how this also
holds for the less familiar 2D case, and then discuss why this makes 3D
delta potential wells unphysical as models of interparticle interactions for
condensed-matter many-body systems. However, both 2D and 3D delta wells can
be ``regularized'' to support a single bound level which in turn renders
them conveniently simple single-parameter interactions, e.g., for modeling
the pair-forming dynamics of quasi-2D superconductors such as the cuprates,
or in 3D of other superconductors and of neutral-fermion superfluids such as
ultra-cold trapped Fermi gases. \\

\noindent PACS numbers: 03.75.Ss; 03.65.-w; 03.65.Ge; 74.78.-w \\
Key words: delta potential wells, bound states, regularization, condensed-matter physics.
\end{abstract}

\centerline{\bf Resumen}
Es bien sabido que un pozo de potencial delta en 1D tiene un solo estado ligado pero que en 3D  tiene  un n\'umero {\it infinito} de estos estados con una energ\'{\i}a de ``amarre" infinita para el nivel m\'as bajo. Aqu\'{\i} mostramos c\'omo esto tambi\'en ocurre para el caso bidimensional que es menos familiar, para luego discutir por que los pozos de potencial delta en 3D no son f\'{\i}sicos como modelos de interacciones entre part\'{\i}culas para sistemas de muchos cuerpos en materia condensada. No obstante, ambos pozos delta en 2D y 3D pueden ser regularizados para soportar un solo  nivel ligado lo cual los convierte convenientemente en interacciones de un solo par\'ametro, por ejemplo, para modelar la din\'amica de formaci\'on de pares en superconductores casi-bidimensionales tales como los cupratos, o en 3D la formaci\'on de pares en otros superconductores y en superfluidos fermi\'onicos neutros tales como los gases de Fermi atrapados ultrafr\'{\i}os.

\pagebreak

\noindent {\bf I. INTRODUCTION}

The study of physical systems in dimensions lower than three has recently
shed its purely academic character and become a real necessity to describe
the properties of novel systems such as{\it \ nanotubes} \cite{nanotubes},
quantum {\it wells}, {\it wires} and {\it dots} \cite{wellanddots,resource},
the Luttinger liquid \cite{resource,Luttinger}, etc. Reduced dimensionality
describes superconducting phenomena in quasi-2D cuprates where pairing
between electrons (or holes) is essential \cite{cuprate}. Whatever the
actual interaction between two electrons (or holes) in a cuprate might
ultimately turn out to be, the attractive delta potential is a conveniently
simple model to visualize and to account for pairing, an indispensable
element for superconductivity and neutral-fermion superfluidity. It
enormously simplifies calculations. Bound states in a delta potential well
in 1D and 3D are usually discussed in textbooks, but not in 2D.{\Large \ }%
Refs. \cite{Albeverio} and \cite{Jackiw} discuss this from a more rigorous
mathematical viewpoint without explicitly solving the Schr\"{o}dinger
equation, e.g., for the bound energy levels.{\Large \ }Here, this gap is
filled by analyzing the 2D time-independent Schr\"{o}dinger equation with a
delta potential well that is then ``regularized'' \cite{Gosdzinsky}\ to
reduce its infinite bound levels to only one. The single-bound-level case
suffices for now since, e.g., the well-known simple Cooper/BCS model
interaction \cite{Cooper} mimicking the attractive electron-phonon
pair-forming mechanism, but requiring two parameters (a strength and a
cutoff) instead of the regularized $\delta $-potential's {\it only one} (a
strength), can also be shown to support a single bound state \cite{IJTP} in
the vacuum or two-body limit. Were it not for the (momentum-space) cutoff
parameter, the Cooper/BCS interaction in coordinate space would also be a $%
\delta $-potential, and indeed becomes such as the cutoff is properly taken
to infinity.

From elementary quantum mechanics we first recall the bound-state energies $%
E<0$ in a potential ``square'' well of depth $V_{0}$ and range $a$, a common
textbook example studied in 1D \cite{Gasiorowicz} and 3D \cite{Park}. In 1D
the ground-state energy $E_{0}$ of a particle of mass $m$ can be expanded
for small $V_{0}a$ as 
\begin{equation}
E_{0}\;{\smash {\mathop{\relbar\joinrel\longrightarrow}\limits_{V_{0}a \to
0}}}\;-\frac{2ma^{2}V_{0}^{2}}{\hbar ^{2}}+O\left( a^{3}V_{0}^{3}\right) .
\label{E01D}
\end{equation}%
Thus, in 1D there is always at least one bound state no matter how shallow
and/or short-ranged the well. Similarly, in 3D for a spherical well, an
expansion of $E_{0}$ in powers of $\eta \equiv V_{0}a^{2}-\hbar ^{2}\pi
^{2}/8m$ $\geq 0$ gives 
\begin{equation}
E_{0}\;{\smash {\mathop{\relbar\joinrel\longrightarrow}\limits_{\eta \to
0^{+}}}}\;\ \frac{m\eta ^{2}}{2\hbar ^{2}a^{2}}+O\left( \eta ^{3}\right) .
\label{E03D}
\end{equation}%
Thus, in contrast to 1D, a minimum critical or threshold value for $%
V_{0}a^{2}$ of $\hbar ^{2}\pi ^{2}/8m$ is needed in 3D\ for the first bound
state to appear. Clearly, both 1D and 3D cases are {\it perturbative}
expansions in an appropriate ``smallness'' parameter, $V_{0}a$ or $\eta $.
As in 1D, a 2D circularly symmetric well of depth $V_{0}$ and radius $a$ 
{\it always} supports a bound state, no matter how shallow and/or
short-ranged the well. However, this instance is {\it non-perturbative }as
it gives $\cite{Landau}$ for the lowest bound-state energy 
\begin{equation}
E_{0}\;{\smash {\mathop{\relbar\joinrel\longrightarrow}\limits_{V_{0}a^2 \to
0}}}\;\ \frac{\hbar ^{2}}{2ma^{2}}\exp \left( -\frac{2\hbar ^{2}}{mV_{0}a^{2}%
}\right)   \label{E02D}
\end{equation}%
which {\it cannot} be expanded in powers of small $V_{0}a^{2}$ since it is
of the form $f(\lambda )=e^{-1/\lambda }%
\mathrel{\mathop{\longrightarrow} \limits_{\lambda
\rightarrow 0}}0,$ i.e., has an {\it essential} singularity at $\lambda =0$.

In this paper we discuss how, just as in the better known 3D case, the 2D
potential well $-v_{0}\delta \left( {\bf r}\right) ,$ $v_{0}>0,$ also
supports an infinite number of bound states with the lowest bound level
being infinitely bound for any fixed $v_{0}$. For an $N\rightarrow \infty $
many-fermion system interacting pairwise via a delta potential, arguments
based on the Rayleigh-Ritz variational principle show that the entire system
in 3D would collapse to infinite binding energy per particle $E/N\rightarrow
-\infty $ and infinite number density $n\equiv N/V\rightarrow \infty $. This
occurs since the lowest two-particle bound level in each $\delta $-well
between pairs is infinitely bound, for any fixed $v_{0}.$ To avoid this
unphysical collapse one generally imagines square wells in 3D (and also in
2D) ``regularized'' \cite{Gosdzinsky} into $\delta $ wells $-v_{0}\delta
\left( {\bf r}\right) $\ that support a {\it single} bound-state, a
procedure leaving an infinitesimally small $v_{0}.$ The remaining $\delta $%
-potential well is particularly useful in condensed-matter theories, e.g.,
of superconductivity \cite{BCS-Bose}\ or neutral-fermion superfluidity \cite%
{Holland01,Griffin02}, where the required Cooper pairing can arise \cite%
{Goodstein} from an arbitrarily weak attractive interaction between the
particles (or holes).

After beginning with a $d$-dimensional expression for the delta potential in
Sec. II, we summarize how bound states emerge in 1D and 3D\ by recalling
textbook results. In Sec. III we analyze in greater detail the less common
2D problem. In Sec. IV we sketch the use of ``regularized'' 2D $\delta $
potential wells for electron (or hole) pairing in quasi-2D cuprates and in
Sec. V we give details for the 2D case. Sec. VI offers conclusions. \newline

\noindent {\bf {II. REVIEW OF DELTA POTENTIAL WELLS IN 1D AND 3D }}

The attractive square potential well in $d$ dimensions 
\begin{equation}
V({\bf r})=-V_{0}\theta (a-r),  \label{Vr}
\end{equation}%
where the Heaviside step function $\theta (x)\equiv \frac{1}{2}[1+\mbox{sgn}%
(x)]$, $a$ is the well range, and $V_{0}\geq 0$ its depth. An attractive
delta potential $-v_{0}\delta ({\bf r})$ $(v_{0}>0)$ can then be constructed
from the double limit 
\begin{equation}
-V_{0}\theta (a-r)\mathrel{\mathop{\overrightarrow{\hspace{1in}}}\limits_{
\begin{array}{c} V_{0}\rightarrow \infty , \, { }a\rightarrow 0 \\ \ni
a^{d}V_{0}= \hbox{const.} \end{array} }}-v_{0}\delta ({\bf r}),
\label{Vtheta}
\end{equation}%
where $v_{0}\equiv c_{d}a^{d}V_{0}$ is a positive constant, with $%
c_{d}\equiv \pi ^{d/2}/\Gamma \left( d/2+1\right) $ as follows on
integrating both sides of (\ref{Vtheta}) over the entire $d$-dimensional
``volume'' and recalling that $\int d^{d}r\delta ({\bf r})=1.$ We seek the
bound-state eigenenergies $E<0$ from the time-independent Schr\"{o}dinger
equation for a particle of mass $m$ in potential (\ref{Vr}), namely 
\begin{equation}
\nabla ^{2}\Psi ({\bf r})-\frac{2m}{\hbar ^{2}}[V({\bf r})+\left| E\right|
]\Psi ({\bf r})=0,  \label{Sch}
\end{equation}%
where $E\equiv -\left| E\right| .$

In 1D the solutions of (\ref{Sch}) (with $r\geq 0$ taken as $\left| x\right|
)$ for $x\neq 0$ are $\Psi (x)=e^{\pm px}.$ These functions have a
discontinuous derivative at $x=a$ in the delta potential limit (\ref{Vtheta}%
) where 
\begin{equation}
\mathrel{\mathop{\lim }\limits_{{\small V}_{0}{\small \rightarrow \infty ,}
\, { }{\small a\rightarrow 0}}}2aV_{0}\equiv v_{0}\quad <\quad \infty ,
\label{V0}
\end{equation}%
and there is always a (single) bound-state energy $E=-mv_{0}^{2}/2\hbar ^{2}$
$\ (v_{0}\neq 0)$ \cite{Gasiorowicz}. Note that in the {\it integral method}
of Ref. \cite{Landau} applicable to shallow wells where $\left| E\right| \ll
\max \left| V(x)\right| ,$ $E$ for 1D would be given as 
\begin{equation}
E=-\frac{m}{2\hbar ^{2}}\left[ \int_{-\infty }^{\infty }dxV(x)\right] ^{2},
\label{E1}
\end{equation}%
which for $V(x)=-v_{0}\delta (x)$ becomes 
\begin{equation}
E=-\frac{m}{2\hbar ^{2}}v_{0}^{2}\left[ \int_{-\infty }^{\infty }dx\delta (x)%
\right] ^{2}=-\frac{mv_{0}^{2}}{2\hbar ^{2}},  \label{E2}
\end{equation}%
which agrees with Ref. \cite{Gasiorowicz} and is consistent with (\ref{E01D}%
). The 1D $\delta $-potential well has proved very convenient in modeling %
\cite{Lieb,LeyKoo}\ self-bound many-fermion systems in 1D, and in
understanding Cooper pairing \cite{Casas91} as well as the BCS theory of
superconductivity \cite{PRB47}.

For the potential (\ref{Vr}) in 3D the particle wave function in spherical
coordinates \cite{Merzbacher} is $\Psi ({\bf r})=R_{l}(r)Y_{lm}(\theta ,\phi
),$ where (Ref. \cite{Arfken}, p. 722) $Y_{lm}\left( \theta ,\phi \right) $
are the spherical harmonics and $R_{l}(r)$ the radial wavefunctions. For $%
0\leq r\leq a$ the finite (or regular) radial solutions are spherical Bessel
functions of the first kind $j_{l}(Kr)$ of order $l$, with $K^{2}\equiv
2m(V_{0}-\left| E\right| )/\hbar ^{2},$ since $j_{l}(Kr)<\infty $ at $r=0.$
For $r\geq a$ the linearly-independent radial solutions are the so-called
modified spherical Bessel functions $k_{l}(kr)$, with $k^{2}\equiv 2m\left|
E\right| /\hbar ^{2},$ where $k_{l}(kr)$ decays exponentially as $%
r\rightarrow \infty .$ The boundary conditions at $r=a$ expressing the
continuity of the radial wave function $R_{l}\left( r\right) $ and of its
first derivative can be combined into the single relation 
\begin{equation}
\left. \frac{dj_{l}(Kr)/dr}{j_{l}({K}r)}\right| _{r=a^{-}}\hspace{0.1in}=%
\hspace{0.1in}\left. \frac{dk_{l}(kr)/dr}{k_{l}(kr)}\right| _{r=a^{+}}.
\label{dJ/J0}
\end{equation}%
Taking $l=0$ and recalling (Ref. \cite{Arfken}, pp. 730, 733) that $%
j_{0}(x)=\sin (x)/x,$ and $k_{0}(x)=e^{-x}/x,$ (\ref{dJ/J0}) gives 
\begin{equation}
{K}\cot ({K}a)=-k,\hspace{1in}(l=0).  \label{Kcot}
\end{equation}%
We can write the $l=0$ bound-state energies $E_{n}=-\hbar ^{2}/2ma^{2}\left(
\pi ^{2}/4+\varepsilon _{n}^{2}\right) ,$ where $\varepsilon _{n}$ are the
dimensionless roots of (\ref{Kcot}), with $n=1,2,...{\bf .}$ The standard
graphical solution \cite{Park}\ of condition (\ref{Kcot}) shows that there
are precisely $n$ bound $l=0$ states whenever the well parameters are such
that \cite{deLltext} 
\begin{equation}
(n-1/2)\pi \quad \leq \quad \left( \frac{2mV_{0}a^{2}}{\hbar ^{2}}\right)
^{1/2}\quad \leq \quad (n+1/2)\pi ;\qquad (n=1,2,...,\infty ).  \label{L=0}
\end{equation}%
Thus, the first bound state ($n=1$) appears when $V_{0}a^{2}\geq \pi
^{2}\hbar ^{2}/8m,$ as was mentioned below (\ref{E03D}), and $n=2$ requires
a deeper well depth $V_{0}$ and/or larger well range, etc.

The 3D delta
potential well $-v_{0}\delta \left( {\bf r}\right) ,$ as defined in (\ref%
{Vtheta}), integrated over all space gives 
\begin{eqnarray}
v_{0} &\equiv &\int d^{3}{\bf r}v_{0}\delta ({\bf r}){\bf =}%
\mathrel{\mathop{\lim }\limits_{{\small V}_{0}{\small \rightarrow \infty ,}
\, { }{\small a\rightarrow 0}}}\int d^{3}{\bf r}V_{0}\theta \left(
a-r\right)   \nonumber \\
&=&\mathrel{\mathop{\lim }\limits_{{\small V}_{0}{\small \rightarrow \infty
,} \, { }{\small a\rightarrow 0}}}\frac{4\pi }{3}V_{0}a^{3}\quad <\quad
\infty .  \label{nu0}
\end{eqnarray}%
Hence, as $V_{0}\rightarrow \infty ,$ $a\rightarrow 0$ the middle term in (%
\ref{L=0}) $(2mV_{0}a^{2}/\hbar ^{2})^{1/2}\equiv (3mv_{0}/2\pi \hbar
^{2}a)^{1/2}\longrightarrow \infty $, so that the number of bound-states $n$
in the 3D delta potential well $-v_{0}\delta ({\bf r})$\ is {\it infinite }%
for any finite fixed strength $v_{0}.$ \\

\noindent{\bf {III. 2D DELTA POTENTIAL WELL}}

This same result holds in 2D but is not as apparent. Here the solutions of (%
\ref{Sch}) are $\Psi ({\bf r})=f(r)e^{i\nu \phi },$ with $\nu =0,1,2,...$
and the angular variable $-\pi \leq \phi \leq \pi .$ For $0\leq r\leq a$ the
radial solutions which are finite at $r=0$ are cylindrical Bessel functions $%
J_{\nu }(Kr)\equiv \sqrt{2Kr/\pi }j_{\nu -1/2}(Kr)$ (Ref. \cite{Arfken}, p.
669) of {\it integer} order $\nu ,$ with $K^{2}\equiv 2m(V_{0}-\left|
E\right| )/\hbar ^{2}.$ For $r>a,$ as linearly-independent solutions one has
the modified Bessel functions $K_{\nu }(kr)$ with $k^{2}\equiv 2m\left|
E\right| /\hbar ^{2},$ which are regular as $r\rightarrow \infty $. The two
boundary conditions at $r=a$ can again be written as a single relation 
\begin{equation}
\left. \frac{dJ_{\nu }({K}r)/dr}{J_{\nu }({K}r)}\right| _{r=a^{-}}\;\ =\;\
\left. \frac{dK_{\nu }(kr)/dr}{K_{\nu }(kr)}\right| _{r=a^{+}}.  \label{dJ/J}
\end{equation}
As we want to ensure against collapse in our many-body system interacting
pairwise with the $\delta $\ potential, it is enough to show this for the
lowest bound level with $\nu =0.$ In this case (\ref{dJ/J}) becomes, since $%
dJ_{0}\left( Kr\right) /dr=-KJ_{1}\left( Kr\right) $ and $dK_{0}\left(
kr\right) /dr=-kK_{1}\left( kr\right) $ (Ref. \cite{Abramowitz}, p. 361 and
376, respectively), 
\begin{equation}
{K}a\frac{J_{1}({K}a)}{J_{0}({K}a)}=ka\frac{K_{1}(ka)}{K_{0}(ka)}.
\label{Ka}
\end{equation}
Since $K_{1}(x)>K_{0}(x)>0$ for all $x$, the rhs of (\ref{Ka}) is always a
positive and increasing function of $ka;$ it is plotted in Fig. 1 for $%
V_{0}/\left| E\right| =300$ (dashed curve). As for the lhs, $J_{0}\left(
x\right) $ oscillates for all $x$ so that it diverges positively whenever $%
J_{0}(x)=0$, then changes sign and thus drives the lhs to $-\infty $ (see
full curve in figure). Clearly, there is always an intersection (bound
state, marked by dots in figure) between two consecutive zeros of $J_{0}(x).$
For a given interval in $ka,$ the closer these poles are, the more
bound-states there will be. {Thus, for any given square well, all of the
allowed bound-states lie inside an interval between }${0}${\ and $k_{\max
}a, $ where $k_{\max }\equiv (2mV_{0}/\hbar ^{2})^{1/2}$. In such an
interval the number $n$ of bound states (zeros) will be $n=$ INT($\alpha
k_{\max }a/\pi $),} with $\alpha \equiv \left( V_{0}/\left| E\right|
-1\right) ^{1/2} $ where the INT($x$) function rounds a number $x$ down to
the nearest integer. Of course, the expression for $n$ is only valid after
the appearance of the first pole. Then for $V_{0}/\left| E\right| =300$ as
in Fig. 1, $n=3$ in the interval between $0$ and $k_{\max }a=0.5.$ In Fig. 2
are shown the bounds for $V_{0}/E=2700$ where there are $n=7$ bound states
in the interval between $0$ and $k_{\max }a=0.4$ as it should be.

%FIGURE 1.

To construct a delta potential well $-v_{0}\delta \left( {\bf r}\right) $ in
2D from the finite-ranged well (\ref{Vr}), and through (\ref{Vtheta}) ensure
that $\int d^{2}{\bf r}\delta \left( {\bf r}\right) =1$, requires that 
\begin{equation}
\mathrel{\mathop{\lim }\limits_{V_{0}\rightarrow \infty , \, { }a\rightarrow
0}}V_{0}\pi a^{2}\equiv v_{0}\quad <\quad \infty .  \label{nu02}
\end{equation}
Thus, as long as $\left| E\right| $ is finite $ka\equiv (\sqrt{2m\left|
E\right| /\hbar ^{2}})a%
\mathrel{\mathop{\longrightarrow
}\limits_{V_{0}\rightarrow \infty , \, { }a\rightarrow 0}}0$ and we can use
(Ref. \cite{Goodstein}, p. 612) $xK_{1}(x)/K_{0}(x)%
\mathrel{\mathop{\longrightarrow
}\limits_{x\ll 1}}-1/\ln x$ for the rhs of (\ref{Ka}). In this case the
number $n$ of bound states for $\delta $-well corresponds again to the
number of zeros of the lhs of (\ref{Ka}) but in the delta limit. Here, from (%
\ref{nu02}) $Ka\equiv ka\left( V_{0}/\left| E\right| -1\right) ^{1/2}$ ${%
\smash{\mathop{\relbar\joinrel\longrightarrow}\limits_{V_{0}\rightarrow
\infty ,a\rightarrow 0}}} 
%\mathrel{\mathop{\longrightarrow }\limits_{V_{0}\rightarrow \infty ,{
%}a\rightarrow 0}}
\sqrt{2mv_{0}/\pi \hbar ^{2}}<\infty $ (not necessarily $\ll 1$). We will
see below that the case $Ka\ll 1$ corresponds to the shallow 2D potential
well of Ref. \cite{Landau}. But even if $Ka$ is not $\ll 1,$ Bessel
functions oscillate for large argument although their period is not
constant. In this latter case (Ref. \cite{Abramowitz}, p. 364) $J_{1}(x)%
\mathrel{\mathop{\longrightarrow }\limits_{x\gg 1}}$ $\sqrt{2/\pi x}\cos
(x-3\pi /4)$ allows locating the zeros of the lhs of (\ref{Ka}) which as $%
V_{0}/\left| E\right| $ is increased approach each other on the $ka${\it \ }%
axis, so that in the delta well limit as $V_{0}/\left| E\right|
\longrightarrow \infty $ the number $n$ of bound-states increases
indefinitely. Moreover, rewriting (\ref{Ka}) as 
\begin{equation}
\alpha x_{n}J_{1}(\alpha x_{n})K_{0}(x_{n})-x_{n}K_{1}(x_{n})J_{0}(\alpha
x_{n})=0,  \label{xn}
\end{equation}
bound states are easily identified from Fig. 3, where the roots of (\ref{xn}%
), say $x_{n}\equiv k_{n}a=\left( \sqrt{2m\left| E_{n}\right| /\hbar ^{2}}\right)a,$ are
seen to form an infinite set as $\alpha \longrightarrow \infty $. Therefore
the 2D delta potential well supports an {\it infinite number of states}, for
any fixed $v_{0}$, precisely as in the 3D case, this being the main
conclusion of the paper. Table 1 shows the first few (numerical) $x_{n}$
roots where $E_{n}\equiv -\hbar ^{2}x_{n}^{2}/2ma^{2},$ for three extreme
values of $V_{0}/\left| E\right| .$

%FIGURE 3.

Applying the integral method of Ref. $\cite{Landau}$ for $\nu =0$ for a
shallow potential well, i.e., $V_{0}\rightarrow 0$ and $\left| E\right| \ll
V_{0},$ one can take both $Ka$ and $ka$ $\rightarrow 0.$ Thus, we can use $%
xK_{1}(x)/K_{0}(x)\mathrel{\mathop{\longrightarrow
}\limits_{x\ll 1}}-1/\ln x$ in the rhs of (\ref{Ka}), and in the lhs of (\ref%
{Ka}) we note that (Ref. \cite{Abramowitz}, p. 360) $J_{\nu }(x)%
\mathrel{\mathop{\longrightarrow
}\limits_{x\ll 1}}x^{\nu }/2^{\nu }\nu !$, with $\nu =0$ and $1,$ so that $%
xJ_{1}(x)/J_{0}(x)\mathrel{\mathop{\longrightarrow
}\limits_{x\ll 1}}$ $x^{2}/2$ . Hence we write (\ref{Ka}) as 
\begin{equation}
-\frac{1}{(Ka)^{2}}\simeq \frac{\ln ka}{2},  \label{1/Ka}
\end{equation}
so that on putting $V_{0}-\left| E\right| \approx V_{0}$ (\ref{1/Ka})
becomes precisely (\ref{E02D}). In fact, for any shallow 2D
circularly-symmetric potential well $V\left( r\right) ,$ the first bound
state in Ref. $\cite{Landau}$ is given by 
\begin{equation}
E\simeq -\frac{\hbar ^{2}}{2ma^{2}}\exp \left( -\hbar ^{2}/m\left|
\int_{0}^{\infty }drrV(r)\right| \right) ,  \label{EE1}
\end{equation}
which for potential (\ref{Vr}) reduces to (\ref{E02D}). This result in the
delta limit of (\ref{nu02}) finally becomes 
\begin{equation}
E\simeq -\frac{\hbar ^{2}}{2ma^{2}}\exp \left( -2\hbar ^{2}\pi
/mv_{0}\right) ,  \label{EvE1}
\end{equation}
where $v_{0}<\infty .$ \\

\noindent {\bf {IV. NEED TO REGULARIZE IN CONDENSED-MATTER SYSTEMS}}

Real condensed matter systems are made of many particles (bosons and/or
fermions) interacting via attractive and/or repulsive forces. Attractive
forces between fermions can form pairs needed for many properties such as
superconductivity in solids or superfluidity in fermion liquids or trapped
atomic fermion gases. However, addressing these problems with a physically
realistic interaction is oftentimes difficult. As in 1D with a ``bare''\ $%
\delta $-potential well, a regularized attractive $\delta $-well prevents
collapse in 3D, provides the required pairs in either 2D or 3D and, of
course, simplifies calculations.

It is easy to imagine a trial wave function whereby, with an attractive bare 
$\delta $-function interfermionic interaction (i.e., {\it before }%
regularization), a 3D $N$-fermion system would have infinitely negative
energy-per-particle (as well as infinite number-density). This is because
the lowest bound level of the two-body $\delta $-well is {\it infinitely deep%
} in 3D, and indeed also in 2D, as was shown in the preceding sections. By
the Rayleigh-Ritz variational principle the expectation energy associated
with the trial wave function is a rigorous upper bound to the exact $N$%
-fermion ground-state energy, and hence produces collapse of the true 3D
ground state of the system as $N\rightarrow \infty $. In this picture, each
particle ``makes its own well''\ but will attract to itself every other
particle, two for each level, to minimize the trial expectation energy. We
thus get an $N$-fermion system as schematically sketched in Fig. 4 (where
the Pauli exclusion principle is explicitly being applied) that {\it %
collapses} as $N\rightarrow \infty .$ To avoid this unphysical collapse in
3D, and at the same time ensure pair formation in either 2D or 3D, one can
``regularize''\ \cite{Gosdzinsky} the 2D and 3D finite interparticle
potential wells so that in the limit the corresponding $\delta $-well
possesses only one ($s$-wave) bound state. This also occurs with the
Cooper/BCS model interaction \cite{IJTP} definable in any $d$\ and with the
bare $\delta $-well in 1D. The single-bound-state $\delta $-well then
ensures that {\it only }pair ``clusters''\ form, in agreement with quantized
magnetic flux experiments in either elemental \cite{classical}\cite%
{classical2} or cuprate superconductors \cite{Gough} in rings where the
smallest flux trapped is found to be $h/2e$ (with $h$ being Planck's
constant and $e$ the electron charge). This contrasts with $h/e$ as London
conjectured just on dimensional grounds, as well as with $h/ne$ ($n=3,4,...$%
) which is{\it \ not} observed in superconductors, as one would expect {\it %
in vacuo }in other many-particle systems with attractions that produce
clusters of any size. The fact that only pair clusters occur with electrons
(or ``holes'') in superconductors is likely associated with clusters forming. \\

\noindent {\bf {V. REGULARIZED 2D DELTA WELL}}

Regularization in either 2D or 3D starts from a finite-range square well and
yields $\delta $-well with an infinitesimally small strength $v_{0}$, as we
now illustrate. To be specific, we concentrate on the regularization of the
2D finite potential wells needed to mimic, in a simple way, the presence of
Cooper pairs in superconductors. We thus seek a two-body square well
interaction such that, in the $\delta $ limit, it possesses only one ($s$%
-wave) bound state. Following Ref. \cite{Gosdzinsky}, for $d=2$ we
substitute (\ref{Vr}) by an effective two-body square-well interaction $%
V_{a}(r)$ which in the limit $a\rightarrow 0^{+}$ becomes $-v_{0}\delta (%
{\bf r})$ with $v_{0}>0$, and is given by 
\begin{equation}
V_{a}(r)=\frac{\hbar ^{2}}{2m^{\ast }}\frac{2}{a^{2}\ln \left|
a/a_{0}\right| }\theta (a-r),\hspace{1in}(a_{0}>a>0)  \label{Vreg2d}
\end{equation}%
where $m^{\ast }=m/2$ is the reduced mass of the pair, $a$ is still the well
range, and $a_{0}$ is an arbitrary parameter that measures the actual
strength of the interaction. Potential (\ref{Vreg2d}) in the delta limit (%
\ref{Vtheta}) gives a $\delta $-well strength $v_{0}\left( a\right) \equiv
-\int d^{2}rV_{a}\left( r\right) =-\pi \hbar ^{2}/m^{\ast }\ln \left|
a/a_{0}\right| >0$, and is thus infinitesimally small as $a\rightarrow 0^{+}$%
. However, this parameter can be eliminated in favor of the binding energy $%
B_{2}\geq 0$ of the single level, which now serves as coupling parameter.
Indeed, using $v_{0}\left( a\right) $ in equation (\ref{EvE1}) for a
shallow-well as in Ref. \cite{Landau} but with{\Large \ }$m^{\ast }${\Large %
\ }instead of{\Large \ }$m$, we obtain for the lowest energy 
\begin{equation}
E=-\frac{\hbar ^{2}}{ma_{0}^{2}}\equiv -B_{2},  \label{B2}
\end{equation}%
where $0\leq B_{2}<\infty $ is the magnitude of the pair binding energy.
This straightforward procedure then guarantees a simple {\it %
finite-lowest-energy} level well---as in 1D, see (9). Once we set the
regularized two-body interaction model, result (\ref{B2}) can be varied as
the coupling describing our superconductor model. One possibility we have
now is to fix $B_{2},$ which fixes the value of $a_{0};$ the second
possibility is to set $B_{2}$ by fixing the parameter $a_{0},$ which can
represent the range of the wave function of the particles. One can also
reduce the infinite number of bound states to only one by shifting the
center of the 2D $\delta $-potential from the origin along the radial axis %
\cite{Wu}; however the topology of this new $\delta $-potential is
unsuitable to simulate real interactions between electrons.

In 3D regularization proceeds similarly \cite{Gosdzinsky}\ as in 2D except
that there is no log term in (\ref{Vreg2d}), and instead of the binding
energy (\ref{B2}) as coupling parameter one employs the $s$-wave scattering
length which is well-defined even if, unlike 2D, the well is too shallow to
support a bound state.

In an $N$-fermion system interacting pairwise via a regularized $\delta $%
-potential, fermions in the Fermi sea\ bind each other by pairs only, as
required by magnetic flux quantization experiments \cite{classical}\cite%
{classical2}\cite{Gough}, see Fig. 5. The $\delta $-potential has been used
extensively in the literature \cite{Miyake,PRB52} to mimic the pair-forming
interfermion interaction in, e.g., quasi-2D cuprate as well as in otherwise
3D superconductors \cite{PRB52,PhysC351}\ and neutral-fermion superfluids %
\cite{Holland01,Griffin02}. \\

%FIGURE 4.

\noindent{\bf {VI. CONCLUSIONS}}

A graphical proof was provided of how a 2D $\delta $-potential well supports
an infinite number of bound-states as does the more familiar 3D $\delta $%
-potential well. Using Rayleigh-Ritz variational-principle upper-bound
arguments, we then illustrated how in 3D the binding energy-per-particle of
an $N$-fermion system must grow indefinitely as the number of fermions
increases. In order to prevent this unphysical collapse in modeling such a
system one can use {\it regularized} $\delta $-potentials in 3D as well as
in 2D that by construction support a single bound state. This provides
useful interfermion interaction models to study 2D and 3D condensed-matter
problems. An appealing motivation for regularized $\delta $-potentials is
that they fit easily within the framework of the time-independent Schr\"{o}%
dinger equation in either coordinate or momentum space. Indeed, solving the
two-body problem in the Fermi sea allows one already to exhibit Cooper
pairing phenomena which is the starting point for any treatment of
superconductivity or neutral-fermion superfluidity. \\

%\pagebreak

\noindent{\bf ACKNOWLEDGMENTS}

We thank M. Fortes and O. Rojo for discussions, and acknowledge
UNAM-DGAPA-PAPIIT (Mexico) grant \# IN106401, and CONACyT (Mexico) grant \#
27828 E, for partial support. MAS thanks Washington University, St. Louis,
MO, USA, for hospitality during a sabbatical year.

\pagebreak

\begin{center}
\bigskip {\LARGE Tables}
\end{center}

%\begin{mathletters}
\[
\begin{tabular}{|c|c|c|c|c|}
\hline
$V_{0}{\bf /}\left| E\right| $ & $x_{1}$ & $x_{2}$ & $x_{3}$ & $x_{4}$ \\ 
\hline
$10$ & $0.3738$ & $1.4216$ & $2.4674$ & $3.5137$ \\ \hline
$10^{3}$ & $0.0218$ & $0.1248$ & $0.2245$ & $0.3240$ \\ \hline
$10^{5}$ & $0.0123$ & $0.0223$ & $0.0323$ & $0.0422$ \\ \hline
\end{tabular}
\]

Table 1: First few roots $x_{n}=(\sqrt{2m\left| E_{n}\right| /\hbar ^{2}})a$
of\ (\ref{xn}) for bound-states $E_{n}$ of\ 2D\ potential\ well\ according
to (\ref{Ka}), for different values of $V_{0}/\left| E\right| .$

\pagebreak 
\begin{figure}[tbh]
\centerline{\epsfig{file=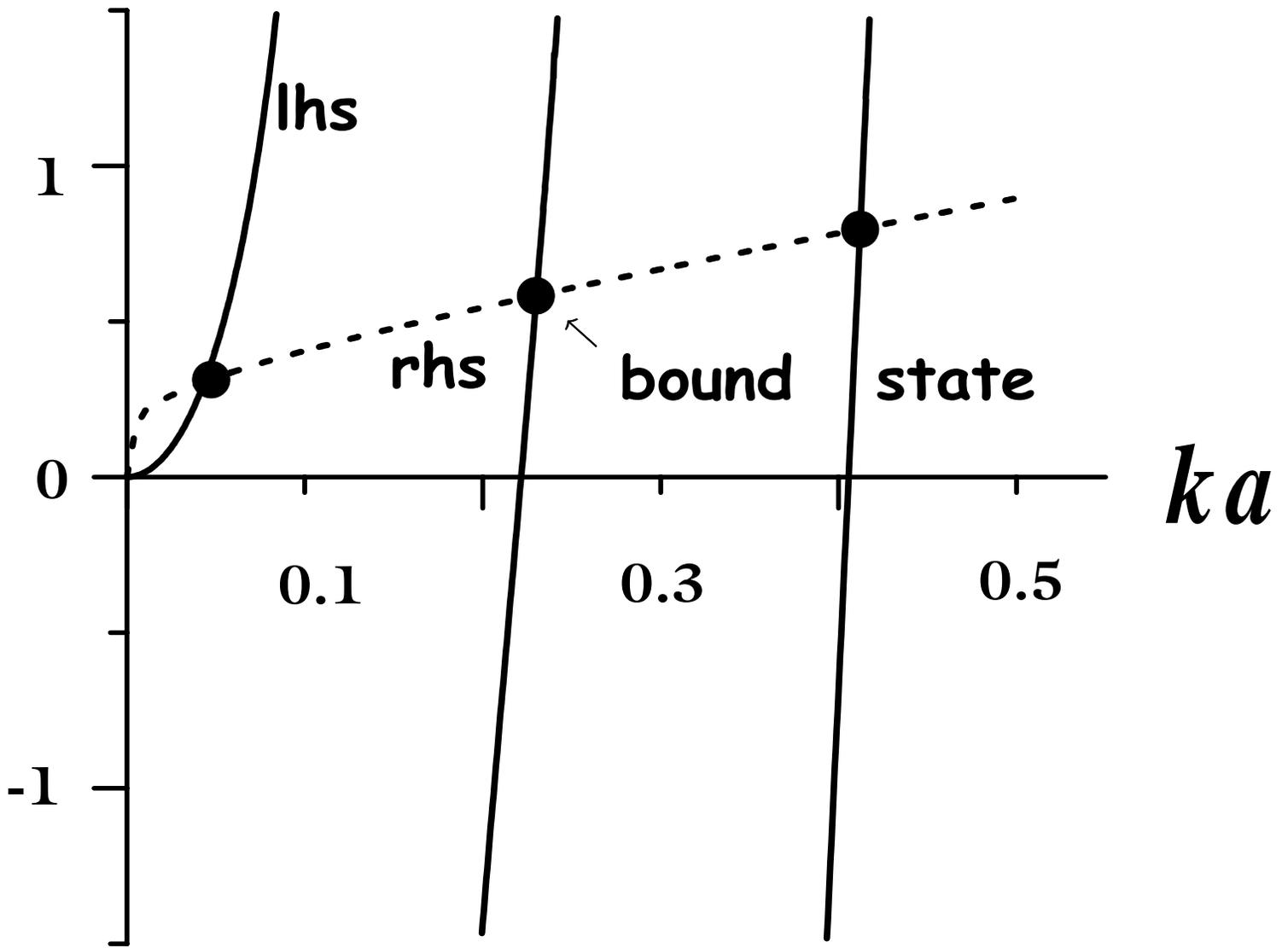,height=7.50in,width=6.0in}} 
%\vspace{-2.0cm}
\caption{Rhs (dashed curve) and lhs (full curve) of (\ref{Ka}) for the 2D
well with $V_{0}/\left| E\right| =300$. Intersections of both curves marked
by dots signal bound states. There is always a bound state between every two
consecutive zeros of $J_{0}(x)$, or poles of the lhs of (\ref{Ka}). }
\end{figure}

%FIGURE 2.
\begin{figure}[tbh]
\centerline{\epsfig{file=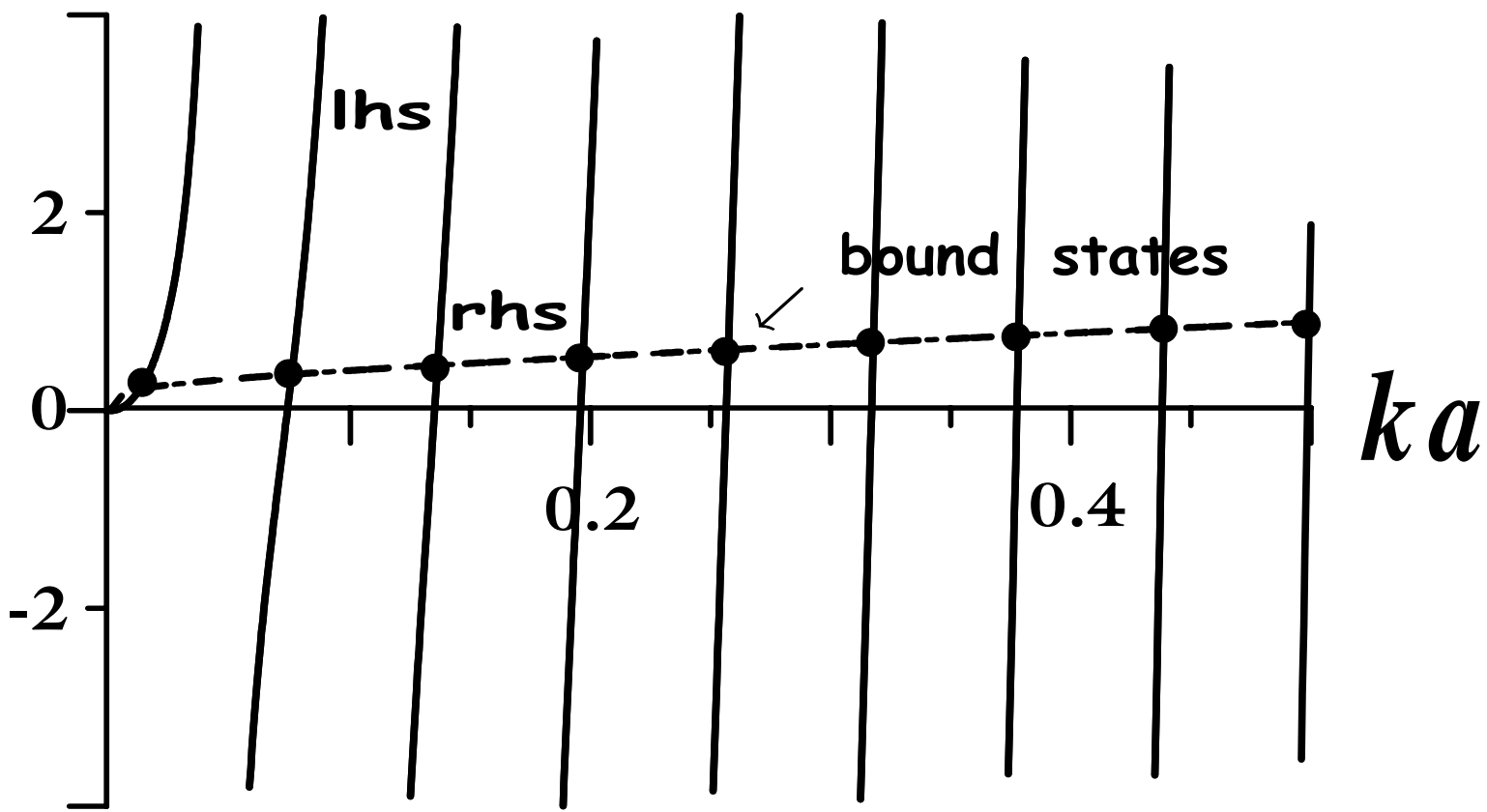,height=7.50in,width=6.0in}} 
%\vspace{-2.0cm}
\caption{Same as Fig. 1 but for $V_{0}/\left| E\right| =2700$\ suggesting
that the number of bound states increases indefinitely as the potential well
approaches the Dirac $\protect\delta $-well limit (\ref{Vtheta}). }
\end{figure}

%FIGURE 3.
\begin{figure}[tbh]
\centerline{\epsfig{file=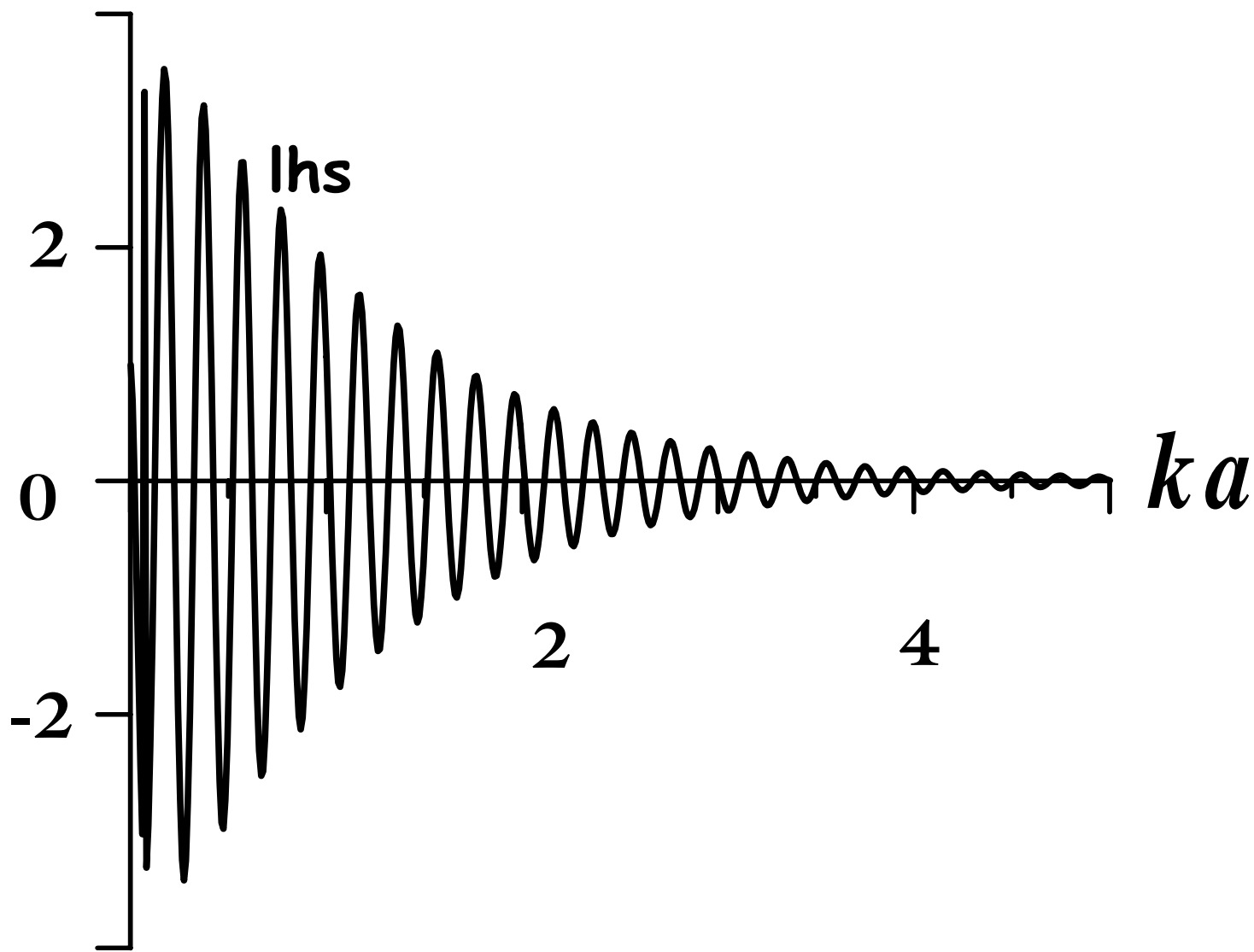,height=7.50in,width=6.0in}} 
%\vspace{-2.0cm}
\caption{The bound states for the 2D $\protect\delta $-well are associated
with the zeros of the lhs of equation (\ref{xn}), plotted on the vertical
axis. For $V_{0}/\left| E\right| =1000$, this graph illustrates the roots
(bound states) of (\ref{xn}) which become an infinite set as we approch to
the $\protect\delta $-well limit. }
\end{figure}

%FIGURE 4.
\begin{figure}[tbh]
\centerline{\epsfig{file=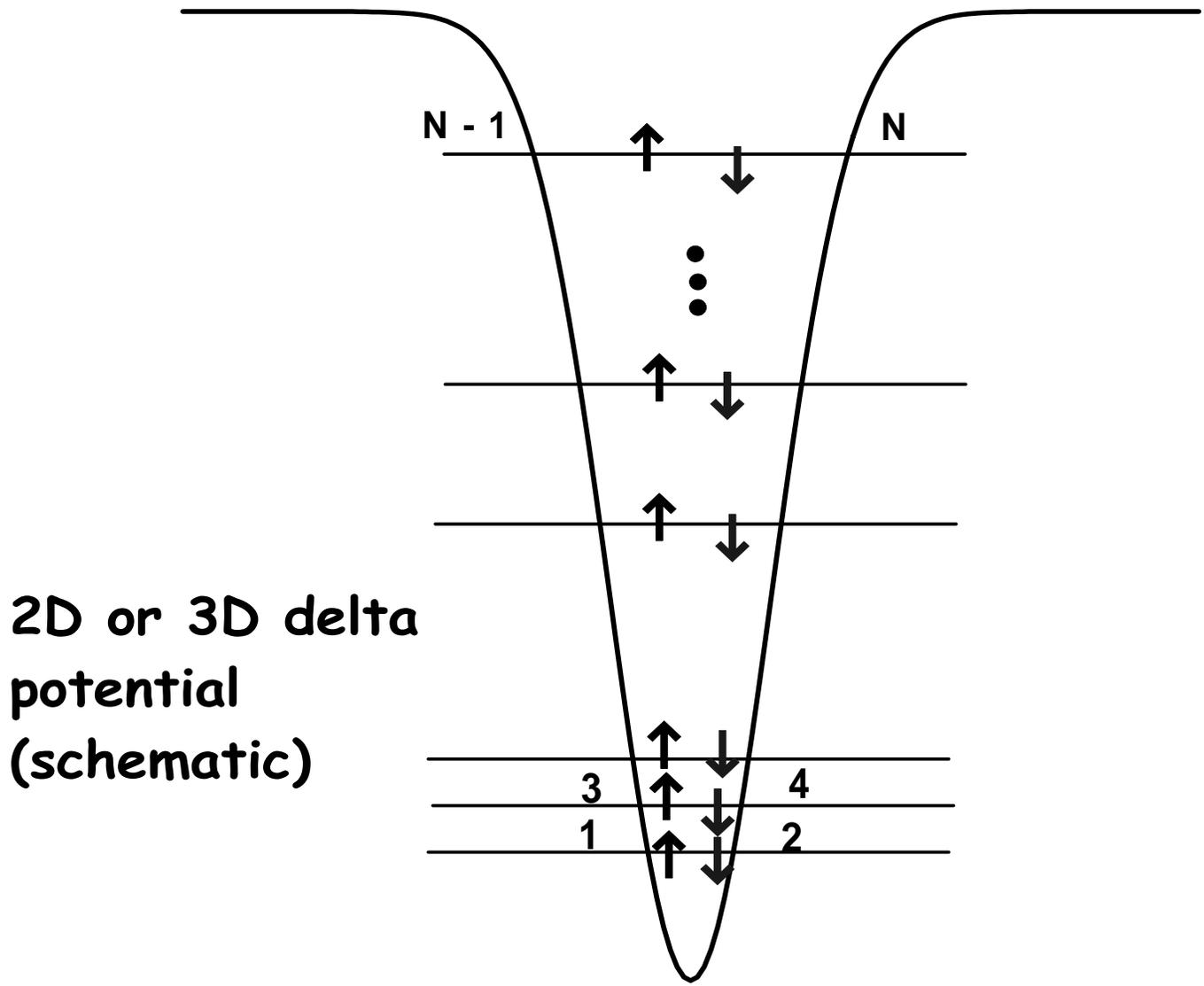,height=7.50in,width=6.0in}} 
%\vspace{-2.0cm}
\caption{An $N$-fermion system with the $\protect\delta $-well pairwise
interactions produces collapse as $N\rightarrow \infty $ in 3D since both
the binding energy per particle and the particle density diverge. }
\end{figure}

%FIGURE 5.
\begin{figure}[tbh]
\centerline{\epsfig{file=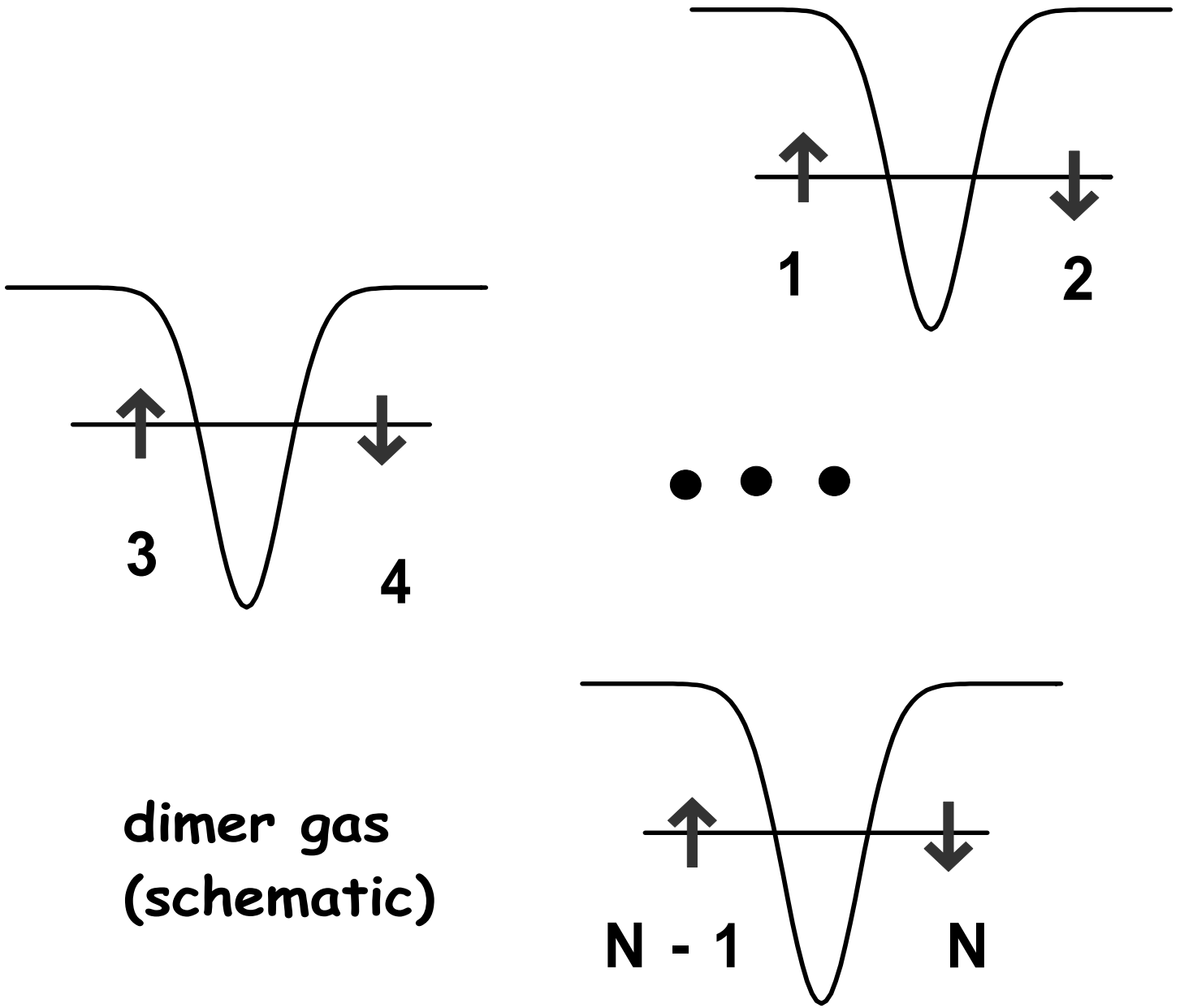,height=7.50in,width=6.0in}} 
%\vspace{-2.0cm}
\caption{A dimer gas formed by single-bound-state regularized $\protect%
\delta $-wells, schematically depicted. In this case, the 3D many-fermion
system will not collapse since the Pauli exclusion principle prevents more
than two particles from being bound in a given well. }
\end{figure}

\end{document}